\documentclass[doublecol]{epl2}

\usepackage{amsmath}
\usepackage{amssymb}
\usepackage{graphicx}
\usepackage{dcolumn}
\usepackage{bm}
\usepackage{color}
\usepackage{xspace}

\title{Excitonic resonances in the 
2D extended Falicov-Kimball model}

\author{Van-Nham Phan\inst{1} \and Holger\ Fehske\inst{1,2} \and Klaus W. Becker\inst{3}}
\shortauthor{V.-N. Phan, H. Fehske, and K. W. Becker}
\institute{\inst{1}Institut f\"ur Physik,
             Ernst-Moritz-Arndt-Universit\"at Greifswald,
             D-17489 Greifswald, Germany\\
\inst{2} School of Physics, University of New South Wales, Kensington 2052, Sydney NSW, Australia\\             
\inst{3}Institut f{\"u}r Theoretische Physik, Technische Universit{\"a}t Dresden, D-01062 Dresden, Germany}


\abstract{
Using  the projector-based renormalization method we investigate the
formation of the excitonic insulator phase in the two-dimensional (2D)
spinless Falicov-Kimball model with dispersive $f$ electrons and address
the existence of excitonic bound states at high temperatures on the 
semiconductor side of the semimetal-semiconductor transition. 
To this end we calculate the imaginary part of the dynamical 
electron-hole pair susceptibility and analyze the wave-vector and energy
dependence of excitonic resonances emerging in the band gap.
We thereby confirm the existence of the exciton insulator
and its exciton environment within a generic two-band 
lattice model with local Coulomb attraction.\vspace*{1pt}}

\begin{document}

\maketitle

\section{Introduction}
\begin{figure}[t]
    \begin{center}
      \includegraphics[angle = -0, width = 0.43\textwidth]{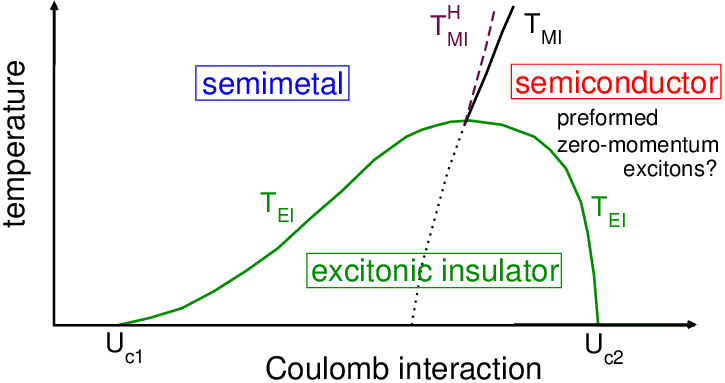}
    \end{center}
\caption{(Color online) Schematic finite-temperature 
phase diagram of the EFKM as obtained from RPA~\cite{IPBBF08} and 
slave-boson~\cite{Br08}  approaches for the particle-hole 
asymmetric case $\varepsilon^f\neq\varepsilon^c$, $0<|t^f|<t^c$. 
At the semimetal-semiconductor transition the ground state of the system
may become unstable with respect to the spontaneous formation and 
condensation of excitons~\cite{Ko68}. 
The strength of the Coulomb interaction determines on which 
side of the metal-insulator transition the system is.
An interesting issue is the possible existence of preformed zero-momentum excitons in the semiconducting region for $T_{EI}<T<T_{MI}$, where $T_{MI}$ denotes the critical temperature for the high-temperature metal-insulator transition. At $T_{MI}^H$ the ``Hartree'' gap opens.}
\label{fig:PD}
\end{figure}

In solids, the Coulomb interaction binds conduction band electrons and valence band holes 
to excitons. Normally, excitonic quasiparticles do not form the ground state but electron-hole
excitations that tend to decay on a very short time scale. At a semimetal-semiconductor
transition,  however, the conventional ground state of the crystal may become unstable
with respect to a spontaneous formation of excitons,  provided the overlap or band gap 
between the valence and conduction bands is small. Then, for low enough temperatures,
these composite bosonic quasiparticles will condense into a macroscopic phase-coherent
quantum state, thereby transforming the semimetallic or 
semiconducting configuration into an insulating one (cf. also fig.~\ref{fig:PD}). This so-called
excitonic insulator (EI) state was theoretically proposed more than four decades 
ago~\cite{Mo61,Ko68};
for recent reviews see~\cite{LEKMSS04,MSGMDCMBTBA10}. 
The EI phase realized below the critical temperature $T_{EI}$ can be perceived either as BCS 
condensate in the semimetal region or as Bose Einstein condensate in the semiconductor region~\cite{BF06}.

The EI state is extremely rare in nature, so far there is no free of doubt realization in any material.
At present, the most promising candidates are the quasi-2D transition-metal dichalgogenide 
$1T$-${\rm TiSe_2}$ and the pressure-sensitive mixed-valence rare-earth chalgogenide 
$\rm TmSe_{0.45}Te_{0.55}$.  In $1T$-${\rm TiSe_2}$,
the excitonic condensate exerts a force on the lattice generating periodic
ionic displacements~\cite{MBCAB11}.  
For $\rm TmSe_{0.45}Te_{0.55}$,
Hall effect, thermal diffusivity and heat conductivity 
data give strong support for  
a Bose condensed state in the pressure range between 5 and 11 
kbar below 20~K ~\cite{NW90}.
The transport anomalies observed 
at higher temperatures, in particular the strange increase of the electrical resistivity in a narrow 
pressure range around 8 kbar might be attributed  to a free-bound state scattering in an 
exciton-rich ``halo'' of an EI~\cite{BF06}.  As a basic prerequisite for the validity of this scenario, the 
existence of free excitons above the EI phase has to be proven, at least for the semiconducting region.

The aim of this paper is to address this issue. The investigation of Falicov-Kimball-type models 
seems to be minimal in this respect.  The original Falicov-Kimball model describes 
localized $f$ electrons and itinerant $c$ electrons interacting by an on-site 
Coulomb interaction~\cite{FK69}.
For our problem, we have to allow for a possible
coherence between conduction band electrons and valence band holes however. This can be
achieved either by including an explicit $c$-$f$ hybridization~\cite{KMM76}
or a finite $f$ bandwidth~\cite{Ba02}.
Indeed, using constrained path Monte Carlo~\cite{BGBL04} 
and mean-field~\cite{Fa08} techniques, 
the 2D Falicov-Kimball model with direct $f$-$f$ particle hopping has been 
shown to exhibit an excitonic ground state 
for intermediate Coulomb couplings provided that the center of 
the $c$ and $f$ bands, $\varepsilon^c$ and $\varepsilon^f$, energetically 
differ.  
Note that around the symmetric case 
$\varepsilon^c=\varepsilon^f$ a charge-density-wave 
phase is energetically more 
stable~\cite{Ba02,BGBL04,Fa08}. 
Recent Hartree-Fock~\cite{SC08}, RPA~\cite{IPBBF08} 
and slave-boson~\cite{Br08} 
studies confirm this finding also for the 3D case. 
In this paper we use the projector-based renormalization method (PRM)~\cite{BHS02,PBF10} to calculate  directly
the excitonic pair susceptibility (up to second order in $U$).  Analyzing the non-trivial frequency- and 
momentum-dependence of $\chi({\bf q},\omega)$ we are able to address the problem of exciton formation
and condensation.

\section{Theoretical approach}
In order to model the generic situation of semiconductors or semimetals with 
short-ranged attractive  Coulomb interaction between conduction 
band ($c$) electrons and valence band ($f$) holes we consider an extended 
version of the Falicov-Kimball Hamiltonian (EFKM),   
\begin{equation}\label{Hami}
\mathcal{H}=\sum_{\mathbf{k}}\bar{\varepsilon}^c_{\mathbf{k}}c^\dagger_{\mathbf{k}}c^{}_{\mathbf{k}}
+\sum_{\mathbf{k}}\bar{\varepsilon}^f_{\mathbf{k}}f^\dagger_{\mathbf{k}}f^{}_{\mathbf{k}}
+U\sum_{i}n^c_in^f_i\,,
\end{equation}
with two dispersive tight-binding bands 
$\bar{\varepsilon}^{c,f}_{\mathbf{k}}=\varepsilon_{}^{c,f}-t^{c,f}_{}
\gamma^{}_{\mathbf k}-\mu\,$.
Here $\varepsilon_{}^{c,f}$ are the on-site energies, $t_{}^{c,f}$
are the nearest-neighbor particle transfer amplitudes, 
$\gamma^{}_{\mathbf k}=2\sum_d^D \cos k_d$ for a D-dimensional 
hypercubic lattice, and $\mu$ denotes the chemical potential.  Accordingly the fermionic operators 
$c^{(\dagger)}_{\mathbf{k}}$ and $f^{(\dagger)}_{\mathbf{k}}$ annihilate 
(create) spinless $c$ and $f$ electrons with momentum $\mathbf{k}$, 
respectively, and $n^c_i$ and $n^f_i$ are the corresponding particle number 
operators for Wannier site~$i$. $U$ parametrizes the local
Hubbard attraction. Note that if the $c$ and $f$ bands are degenerate, $\varepsilon^{c}=\varepsilon^{f}$ and
$t^{c}=t^{f}$, the EFKM reduces to the standard Hubbard model~\cite{Hu63}, 
whereas
for $t^{f}=0$ the genuine FKM arises~\cite{FK69}.
In the latter case the local $f$ 
electron number is strictly conserved~\cite{SC08}. In what follows, 
we study the half-filled band case, with 
total electron density 
$\langle n\rangle =\langle n^c_i\rangle+\langle n^f_i\rangle=1$.
Moreover we consider a direct band gap situation 
with the maximum (minimum) of the $c$ ($f$) band 
dispersion located at $(\pi,\pi)$, i.e. $t^f<0$.  
Without loss of generality the $c$ electrons are considered 
to be `light'  while the $f$ electrons are `heavy', i.e. $|t^f|<1$, 
where the $c$ electron 
hopping integral is taken to be the unit of energy, 
$t^c=1$, and $\varepsilon^c=0$. 

The projector-based renormalization approach starts from the decomposition 
of the many-particle Hamiltonian~(\ref{Hami}) into an ``unperturbed'' part  
$\mathcal{H}_0$ ($c$ and $f$ electron band terms)  and 
into a `perturbation' $\mathcal{H}_1$ (Coulomb interaction term), 
where the unperturbed part $\mathcal{H}_0$ clearly is solvable. 
Then, in general, $\mathcal{H}_1$ accounts for all transitions between the eigenstates of $\mathcal{H}_0$ with nonzero transition energies. Using a series of unitary transformations to integrate out the perturbation ${\mathcal H}_1$ (for details see Ref.~\cite{BHS02}) one arrives at a  final Hamiltonian which is diagonal or at least quasi-diagonal. To evaluate the expectation value $\langle {\mathcal A}\rangle$ of any operator ${\mathcal A}$ also the operator has to be transformed by the same unitary transformation. One of the main advantages of the method is to find broken symmetry solutions of phase transitions~\cite{SB09}.
Note that for practical applications the unitary transformations should best be done in small steps in energy.  Therefore, the evaluation of the transformation in each small step can be restricted to low orders in 
${\cal H}_1$. This procedure usually limits the validity of the renormalization approach to parameters values of $\mathcal H_1$ which are of the order of those of
$\mathcal H_0$. In the present case, good agreement with exact 
results is expected for $U$ values smaller than $t^{c,f}$ or 
$|\varepsilon^f_{\mathbf k} -\varepsilon^c_{\mathbf k}|$ (cf. 
eq.~(\ref{Hartreeshift}) below).  
\section{Excitonic insulator phase}
In a first step, let us address the formation of the long-range 
ordered EI state in the 2D EFKM. To this end we look for a non-vanishing excitonic expectation value $\langle c^\dagger f^{}\rangle$, indicating a spontaneous symmetry breaking 
due the pairing of $c$ electrons ($t^c>0$) with $f$ holes ($t^f <0$). 
Employing the normal-ordered representation of fermionic operators $(: \ldots :)$, the Hamiltonian (\ref{Hami}) reads $\mathcal H = \mathcal H_0 +\mathcal H_1$
with
\begin{eqnarray}\label{HFourierNO}
\mathcal{H}_0&=&\sum_{\mathbf{k}}\varepsilon^c_{\mathbf{k}}:c^\dagger_{\mathbf{k}}c^{}_{\mathbf{k}}:
+\sum_{\mathbf{k}}\varepsilon^f_{\mathbf{k}}:f^\dagger_{\mathbf{k}}f^{}_{\mathbf{k}}:  \nonumber \\  
&&-\sum_{\mathbf{k}}\left(\mit{\Delta}:f^\dagger_{\mathbf{k}}c^{}_{\mathbf{k}}:+\textrm{H.c.}\right) \\
\mathcal H_1 &=&
\frac{U}{N}\sum_{\mathbf{k}_1\mathbf{k}_2\mathbf{k}_3}
:a^{}_{\mathbf{k}_1\mathbf{k}_2\mathbf{k}_3}:\,,\nonumber
\end{eqnarray} 
where
\begin{equation}
 \mit{\Delta}  = \frac{U}{N}\sum_{\mathbf{k}}{d}^{}_{\mathbf{k}} \,.
\label{Delta}
\end{equation}
Here ${d}^{}_{\mathbf{k}}=\langle c^\dagger_{\mathbf{k}}f^{}_{\mathbf{k}}\rangle$ 
plays the role of the EI order parameter, and
$a^{}_{\mathbf{k}_1\mathbf{k}_2\mathbf{k}_3}=c^\dagger_{\mathbf{k}_1}
c^{}_{\mathbf{k}_2}
f^\dagger_{\mathbf{k}_3}f^{}_{\mathbf{k}_1+\mathbf{k}_3-\mathbf{k}_2}$.
Note that in $\mathcal H_0$, the on-site energies are  
shifted by a Hartree term, 
\begin{equation}
\varepsilon^{c,f}_{\mathbf{k}}= \bar{\varepsilon}^{c,f}_{\mathbf{k}}
+U\langle n^{f,c}\rangle\,, 
\label{Hartreeshift}
\end{equation}
where $\langle n^c\rangle=\tfrac{1}{N}
\sum_{\mathbf{k}} \langle c^\dagger_{\mathbf{k}}c^{}_{\mathbf{k}}\rangle$ 
and $\langle n^f\rangle=\tfrac{1}{N}
\sum_{\mathbf{k}} \langle f^\dagger_{\mathbf{k}}f^{}_{\mathbf{k}}\rangle$  
are the mean particle number densities of $c$ and $f$ electrons for  
a system  with $N$ lattice sites. Thus, $\mathcal H_0$ alone corresponds to the Hartree Hamiltonian. Vice versa, only fluctuation operators from the $U$ term 
contribute to ${\mathcal H}_1$. Evaluating the expectation values 
$\langle \ldots \rangle$, the temperature enters into the calculation
via the Fermi function, see~\cite{PBF10}.   

Following the procedure of the PRM approach~\cite{PBF10} 
by integrating out all transitions due to $\mathcal H_1$,  the Hamiltonian 
$\cal H$  can be transformed to a fully renormalized Hamiltonian
\begin{eqnarray}
\label{ReH}
\tilde{\mathcal{H}}&=&
 \sum_{\mathbf{k}}\tilde{\varepsilon}^c_{\mathbf{k}}:c^\dagger_{\mathbf{k}}c^{}_{\mathbf{k}}:
+\sum_{\mathbf{k}}\tilde{\varepsilon}^f_{\mathbf{k}}:f^\dagger_{\mathbf{k}}f^{}_{\mathbf{k}}:\nonumber\\
&&+\sum_{\mathbf{k}}(\tilde{\Delta}^{}_{\mathbf{k}}:f^\dagger_{\mathbf{k}}c^{}_{\mathbf{k}}:+\textrm{H.c.}),
\end{eqnarray}
with modified parameters 
$\tilde{\varepsilon}^c_{\mathbf{k}}$, $\tilde{\varepsilon}^f_{\mathbf{k}}$, 
and $\tilde \Delta_{\mathbf k}$. Note that they take 
important correlation effects into account, which enter from the
elimination procedure.  

Self-evidently the single particle operators have to be transformed in order to evaluate expectation values, i.e. 
\begin{eqnarray}
:\tilde{c}^\dagger_{\mathbf{k}}:\!\!&=&\!\!\tilde{x}^{}_{\mathbf{k}}:c^\dagger_{\mathbf{k}}:+\frac{U}{N}\sum_{\mathbf{k}_1\mathbf{k}_2}
\tilde{y}^{}_{\mathbf{k}_1\mathbf{k}\mathbf{k}_2}:c^\dagger_{\mathbf{k}_1}
f^\dagger_{\mathbf{k}_2}f^{}_{\mathbf{k}_1+\mathbf{k}_2-\mathbf{k}}:\\
:\tilde{f}^\dagger_{\mathbf{k}}:\!\!&=&\!\!\tilde{x}'_{\mathbf{k}}:f^\dagger_{\mathbf{k}}:\nonumber+\frac{U}{N}\sum_{\mathbf{k}_1\mathbf{k}_2}
\tilde{y}'_{\mathbf{k}_1\mathbf{k}_2,\mathbf{k}-\mathbf{k}_1+\mathbf{k}_2}\nonumber\\&&\hspace*{3cm}\times
:c^\dagger_{\mathbf{k}_1}c^{}_{\mathbf{k}_2}f^\dagger_{\mathbf{k}-\mathbf{k}_1+\mathbf{k}_2}:\,,
\label{ren_eq}
\end{eqnarray}
where ${\tilde x}_{\mathbf k}$, ${\tilde y}_{\mathbf k}$, $\cdots$ are also renormalized parameters.
In the  PRM the renormalization results from integrating difference equations with initial conditions taken over from the original Hamiltonian \eqref{HFourierNO}
and the original single particle operators: 
$\varepsilon^{c(f)}_{\mathbf{k},\Lambda}=\varepsilon^{c(f)}_{\mathbf{k}}$, 
 $\Delta^{}_{\mathbf{k},\Lambda}= 0^+$,
$x^{}_{\mathbf{k},\Lambda}(x'_{\mathbf{k},\Lambda})=1$, and $y_{\mathbf{k}_1\mathbf{k}_2\mathbf{k}_3,\Lambda}(y'_{\mathbf{k}_1\mathbf{k}_2\mathbf{k}_3,\Lambda})=0$. 

The final Hamiltonian~\eqref{ReH} can be diagonalized by a 
Bogoliubov transformation  
\begin{eqnarray}\label{HFanoRe}
{\tilde{\mathcal H}}
=\sum_{{\mathbf k}}E^c_{{\mathbf k}}:c^\dagger_{{\mathbf k}}c^{}_{{\mathbf k}}:
+\sum_{{\mathbf k}}E^f_{{\mathbf k}}:f^\dagger_{{\mathbf k}}f^{}_{{\mathbf k}}:
+\tilde{E}\,,
\end{eqnarray}
where the quasiparticle energies are given by 
\begin{equation}
\label{qpe_c}
E^{c/f}_{\mathbf k}=\frac{\tilde{\varepsilon}^c_{\mathbf k}+\tilde{\varepsilon}^f_{\mathbf k}}{2}
\mp\frac{\textrm{sgn}(\tilde{\varepsilon}^f_{\mathbf k}-\tilde{\varepsilon}^c_{\mathbf k})}
{2}W_{\mathbf k}
\end{equation}
with 
$W_{\mathbf k}=[(\tilde{\varepsilon}^c_{\mathbf k}-\tilde{\varepsilon}^f_{\mathbf k})^2
+4|\tilde{\Delta}^{}_{\mathbf k}|^2]^{1/2}$.

A finite $\tilde{\Delta}^{}_{\mathbf k}$ signals $c$--$f$ electron coherence connected with  
a band gap that stabilizes the EI phase. Outside the EI phase, where $\tilde \Delta_{\mathbf k}=0$, a band gap may also exist, provided that 
\begin{equation}
E_g^{(0)}= E_{\mathbf 0}^c  - E_{\mathbf 0}^f = \tilde{\varepsilon}^c_{\mathbf 0}-\tilde{\varepsilon}^f_{\mathbf 0}\, .
\label{Eg0}
\end{equation} 
Therefore $E_g^{(0)}<0$ ($E_g^{(0)}>0$) may be taken as indication that the system is in the 
semimetallic (semiconducting) regime (cf. fig.~\ref{fig:PD}).  Let us emphasize that $E_g^{(0)}$ contains the fully renormalized quasiparticle energies $\tilde{\varepsilon}^{c,f}$, not just the Hartree energies $\varepsilon^{c,f}$ given by~eq.~\eqref{Hartreeshift}.

 For the 2D tight-binding band case studied in this paper, 
we work on a discrete set of $N=24\times 24$ lattice sites and determine all quantities with a
relative error of less than $10^{-5}$. 

\subsection{Semimetallic region} 
\begin{figure}[t]
    \begin{center}
      \includegraphics[angle = -0, width = 0.46\textwidth]{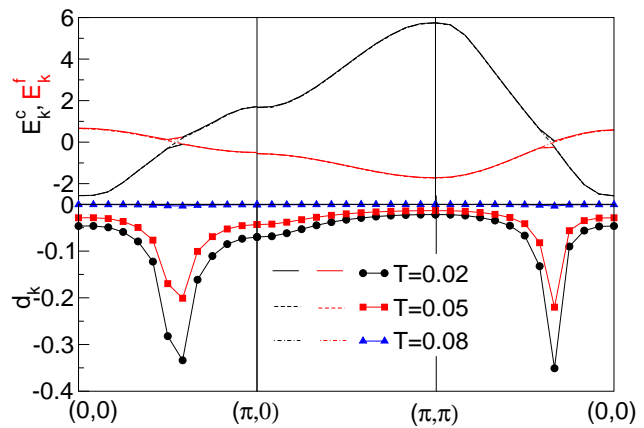}
    \end{center}
\caption{(Color online) Quasiparticle dispersions 
$E^c_{\mathbf k}$ (black lines), $E^f_{\mathbf k}$ (red lines) [upper panel], 
and (b) ``order-parameter'' functions  
$d_{\mathbf k}$ [lower panel] along the major axes of the 
square lattice Brillouin zone. Results are calculated for 
$\varepsilon^f=-1.0$, $t^f=-0.3$, and $U=2.0$ at various 
temperatures.}
\label{fig:Ek}
\end{figure}

Figure~\ref{fig:Ek} (upper panel) shows the renormalized quasiparticle bands $E_{\mathbf k}^c$ and $E_{\mathbf k}^f$ along the high-symmetry axes of the 2D Brillouin zone, for  $\varepsilon^f=-1$, $t^f=-0.3$, and $U=2$. In this case,
both bands overlap ($E_g^{(0)} <0 $) leading to a large Fermi surface,  
where both types of quasiparticles participate. At low temperatures a gap opens at the Fermi surface 
due to the formation of an excitonic insulating state.  
Such an EI state has been viewed before as a BCS condensate of loosely bound electron-hole
pairs~\cite{BF06}. Increasing the temperature above some critical temperature $T_{EI}$ the gap vanishes. 
At this temperature  the EI-semimetal transition takes place. The lower panel of  fig.~\ref{fig:Ek} 
displays the order parameter function $d_{\mathbf k}$ as a function of $\mathbf k$. 
For low temperatures and $\bf k$ close to the Fermi surface, where both quasiparticle bands overlap,
$d_{\mathbf k}$ is strongly peaked. Otherwise $d_{\mathbf k}$ is a rather smooth function of $\mathbf k$.
As a matter of course, increasing $T$ above $T_{EI}$, the order parameter function $d_{\mathbf k}$ vanishes.

\begin{figure}[h!]
    \begin{center}
      \includegraphics[angle = -0, width = 0.46\textwidth]{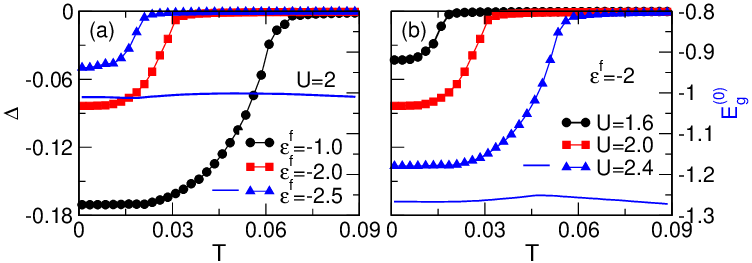}
    \end{center}
\caption{(Color online) 
Temperature dependence of the EI order parameter $\Delta$ 
in (a) for various $\varepsilon^f$ at $U=2$ and in (b) for various  $U$ at 
$\varepsilon^f=-2.0$ where $t^f=-0.3$. The solid lines give the
variation of the gap $E^{(0)}_g$~\eqref{Eg0} (note the scale 
on right-hand ordinate.}
\label{fig:DT}
\end{figure}

In fig.~\ref{fig:DT} [panel (a)] the EI order parameter $\Delta$ [eq.~\eqref{Delta}] is shown 
as a function of temperature for various values of $\varepsilon^f$ at $U=2$.  
Clearly seen is  the formation of an EI state with nonzero $\Delta$ at low temperatures. The EI  
state is  weakened by lowering $\varepsilon^f$ since the overlap of $c$ and $f$ electron bands 
is reduced in this case. In panel (b), the temperature-dependence of $\Delta$
is illustrated for various values of the Coulomb interaction $U$. Similar as before, 
the formation of an EI state is observed for low $T$. The EI region 
is decimated by 
lowering the $c$ electron $f$ hole attraction. 
One can assure oneself that the EI phase only appears in between some 
lower critical value $U_{c1}$ and some upper critical value $U_{c2}$ (on the semiconductor side, see below).   
In both panels of fig.~\ref{fig:DT}, the solid blue lines give
 the variation of the negative quasiparticle gap $E_g^{(0)}$ with $T$, where 
$\varepsilon^f =-2.5$ in panel (a) and $U= 2.4$ in  (b). In either case, the 
variation of $E_g^{(0)}$ for  small $T$ towards lower values goes along 
with the formation of the EI state.  For $U < U_{c1}$ the bare band splitting
$\varepsilon^f-\varepsilon^c$ is somewhat reduced but $E^{(0)}_g$ is still
negative, so we end up with a semimetallic situation.  

\subsection{Semiconducting region} 
\begin{figure}[tb!]
    \begin{center}
      \includegraphics[angle = -0, width = 0.48\textwidth]{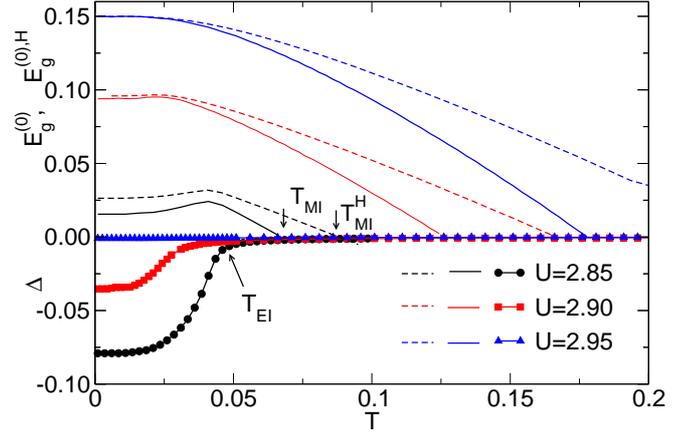}
    \end{center}
\caption{(Color online) EI order parameter $\Delta$ (symbols) and energy gaps 
$E_g^{(0)}$ (solid lines),  $E_g^{(0),H}$ (dashed lines) as functions of temperature for several values of the  
Coulomb interaction $U$ where $\varepsilon^f=-2.4$, $t^f=-0.3$. 
$T_{EI}$ and $T_{MI}$ denote the critical temperatures for the 
EI transition and the metal-insulator transition, respectively.
Note that directly opposed to fig.~\ref{fig:DT} the order parameter 
$\Delta$ decreases with increasing $U$ in the semiconducting regime because 
the effective band overlap (necessary to establish $c$-$f$ electron 
coherence) is now reduced on account of the Hartree term.} 
\label{fig:ED_T}
\end{figure}

We now discuss the possible appearance of an EI state 
on the semiconductor-side of the schematic phase diagram shown in fig.~\ref{fig:PD}.  
Figure~\ref{fig:ED_T} displays the temperature dependence of the quasiparticle  gap $E_g^{(0)}$ and the order parameter $\Delta$ for some larger values of $U$ than before. The order parameter  $\Delta$ is finite  and negative at low $T$   
which again signals the existence of an EI phase. The EI phase in this region was interpreted as a BEC of preformed 
tightly bound excitons~\cite{BF06}.  
For $T$ above the critical temperature $T_{EI}$ the order parameter vanishes and no broken-symmetry state exists as in the semimetallic case. Note that $E_g^{(0)}$ 
is now positive at low temperatures, which indicates that we have a situation with a semiconductor-like band structure 
at least up to some critical temperature $T_{MI}$, where the band gap closes. Obviously, at $T_{MI}$,  
a semiconductor-semimetal metal-insulator transition occurs. 
We further note that due to the inclusion of correlation effects  
the PRM metal-insulator transition temperature $T_{MI}$ strongly deviates from 
$T_{MI}^H$ obtained by using the Hartree-shifted bare 
energies $\varepsilon_{\bf k}^{c,f}$ only, 
where $E_g^{(0),H}= E_{\mathbf 0}^c  - E_{\mathbf 0}^f = \varepsilon^c_{\mathbf 0}-\varepsilon^f_{\mathbf 0}\, $ marks the corresponding Hartree band gap. 
The essential question whether exitonic bound states might possibly exist in the temperature region $T_{EI} < T < T_{MI}$ will be investigated below. In this connection,
in ref.~\cite{BF06} the authors proposed a so-called ``halo phase'',   where individual valence bond holes, conduction band electrons, and bound (but uncondensed) electron-hole pairs (excitons) should coexist.

\section{Excitonic resonances} To address the possible formation of excitonic bound states above $T_{EI}$, 
we analyze the frequency and momentum dependence of the 
dynamical excitonic susceptibility
\begin{equation}
\chi({\mathbf q},\omega)=\langle\langle b^{}_{\mathbf{q}};b^{\dagger}_{\mathbf{q}} \rangle\rangle^{}(\omega)\,,
\label{suszi1}
\end{equation}
where the symbol $\langle\langle \ldots\rangle\rangle$ 
denotes the retarded Green's function, and the creation operator of an electron-hole excitation with momentum $\mathbf q$ is defined by
$b^{\dagger}_{\mathbf{q}}=\frac{1}{\sqrt{N}}\sum_{\mathbf k}c^\dagger_{\mathbf{k}+\mathbf{q}}f^{}_{\mathbf k}$.
Using the unitary invariance of expectation values, 
Eq.~\eqref{suszi1} 
can be rewritten as 
\begin{equation}
\chi({\mathbf q},\omega)=\frac{1}{N}\sum_{\mathbf{k}\mathbf{k}'}\langle\langle \tilde{f}^{\dagger}_{\mathbf k}\tilde{c}^{}_{\mathbf{k}+\mathbf{q}};\tilde{c}^\dagger_{\mathbf{k}'+\mathbf{q}}\tilde{f}^{}_{\mathbf{k}'}\rangle\rangle^{}_{\tilde{\mathcal H}}(\omega)\,.
\label{suszi2}
\end{equation} 
Here the two-particle Green's function on the right-hand side is formed with $\tilde{\mathcal H}$ and the quantities 
with tilde symbols are the fully renormalized operators. Taking into account that the 
EI order parameter vanishes  for $T>T_{EI}$ ($\tilde \Delta_{\mathbf{k}}=0$ $\forall  \mathbf{k}$), 
we obtain up to order ${\cal O}(U^2)$ 
\begin{align}
\label{chi1}
\chi(\mathbf{q}&,\omega)=\frac{1}{N}\sum_{\mathbf{k}}\frac{\Gamma^{0}_{\mathbf{kq}}}{\omega-\omega^{}_{\mathbf k}(\mathbf q) + i\eta}\nonumber\\
+&\frac{1}{N^3}\sum_{\mathbf{k}\mathbf{k}_1\mathbf{k}_2}
\Big[\frac{\Gamma^{1}_{\mathbf{kq}\mathbf{k}_1\mathbf{k}_2}}{\omega-E^{(1)}_{\mathbf{kq}\mathbf{k}_1\mathbf{k}_2}+i\eta}-\frac{\Gamma^{2}_{\mathbf{kq}\mathbf{k}_1\mathbf{k}_2}}{\omega-E^{(2)}_{\mathbf{kq}\mathbf{k}_1\mathbf{k}_2}+i\eta}\Big]\
\end{align}
with
\begin{eqnarray}\label{phe}
\omega^{}_{\mathbf k}(\mathbf q)&=&E^c_{\mathbf{k}+\mathbf{q}}-E^f_{{\mathbf k}}\,,
\\\label{er1}
E^{(1)}_{\mathbf{kq}\mathbf{k}_1\mathbf{k}_2}&=&E^c_{\mathbf{k}_1}-E^f_{\mathbf{k}}
-E^f_{{\mathbf k}_1+\mathbf{k}_2-\mathbf{k}-\mathbf{q}}+E^f_{\mathbf{k}_2}\,,\\\label{er2}
E^{(2)}_{\mathbf{kq}\mathbf{k}_1\mathbf{k}_2}&=&E^c_{\mathbf{k}+{\mathbf q}}-E^c_{\mathbf{k}_1}+E^c_{{\mathbf k}_2}-E^f_{\mathbf{k}-\mathbf{k}_1+\mathbf{k}_2}\,,
\end{eqnarray}
and $\eta=0^+$. The coefficients $\Gamma^i$ are given by
\begin{align}\label{GammaX}
\Gamma^0_{\mathbf{kq}}=&\Big\{|\tilde{x}'_{\mathbf{k}}\tilde{x}^{}_{\mathbf{k}+\mathbf{q}}|^2\nonumber\\
-&\frac{2U}{N}\sum_{\mathbf{k}_1}\tilde{x}'_{\mathbf{k}}\tilde{x}^{}_{\mathbf{k}+\mathbf{q}}(\tilde{x}_{\mathbf{k}_1+\mathbf{q}}\tilde{y}'_{\mathbf{k}_1+\mathbf{q},\mathbf{k}+\mathbf{q},\mathbf{k}}\langle \tilde{n}^c_{\mathbf{k}_1+\mathbf{q}}\rangle\nonumber\\
&\quad\quad\quad+\tilde{x}'_{\mathbf{k}_1}\tilde{y}_{\mathbf{k}+\mathbf{q},\mathbf{k}_1+\mathbf{q},\mathbf{k}_1}\langle \tilde{n}^f_{\mathbf{k}_1}\rangle) \nonumber\\
+&\frac{U^2}{N^2}\sum_{\mathbf{k}_1\mathbf{k}_2}\Big[2\tilde{x}'_{\mathbf{k}}\tilde{x}^{}_{\mathbf{k}+\mathbf{q}}\tilde{y}_{\mathbf{k}_2,\mathbf{k}_1+\mathbf{q},\mathbf{k}_1-\mathbf{k}_2+\mathbf{k}+\mathbf{q}}\nonumber\\
&\times\tilde{y}'_{\mathbf{k}_2,\mathbf{k}+\mathbf{q},\mathbf{k}_1-\mathbf{k}_2+\mathbf{k}+\mathbf{q}}\langle \tilde{n}^c_{\mathbf{k}_2}\rangle\langle \tilde{n}^f_{\mathbf{k}_1-\mathbf{k}_2+\mathbf{k}+\mathbf{q}}\rangle\nonumber\\
&+2\tilde{x}'_{\mathbf{k}_1}\tilde{x}^{}_{\mathbf{k}_2+\mathbf{q}}\tilde{y}_{\mathbf{k}_2,\mathbf{k}_1+\mathbf{q},\mathbf{k}_1}\tilde{y}'_{\mathbf{k}_2+\mathbf{q},\mathbf{k}+\mathbf{q},\mathbf{k}}\langle \tilde{n}^f_{\mathbf{k}_1}\rangle\langle \tilde{n}^c_{\mathbf{k}+\mathbf{q}}\rangle\nonumber\\
&+\tilde{x}_{\mathbf{k}_1+\mathbf{q}}\tilde{x}^{}_{\mathbf{k}_2+\mathbf{q}}
\tilde{y}'_{\mathbf{k}_2+\mathbf{q},\mathbf{k}+\mathbf{q},\mathbf{k}}\tilde{y}'_{\mathbf{k}_1+\mathbf{q},\mathbf{k}+\mathbf{q},\mathbf{k}}\nonumber\\
&\times\langle \tilde{n}^c_{\mathbf{k}_1+\mathbf{q}}\rangle\langle \tilde{n}^c_{\mathbf{k}_2+\mathbf{q}}\rangle\nonumber\\
&+\tilde{x}'_{\mathbf{k}_1}\tilde{x}'^{}_{\mathbf{k}_2}
\tilde{y}_{\mathbf{k}+\mathbf{q},\mathbf{k}_2+\mathbf{q},\mathbf{k}_2}\tilde{y}_{\mathbf{k}+\mathbf{q},\mathbf{k}_1+\mathbf{q},\mathbf{k}_1}\langle \tilde{n}^f_{\mathbf{k}_1}\rangle\langle \tilde{n}^f_{\mathbf{k}_2}\rangle\Big]\Big\}\nonumber\\
&\times(\langle \tilde{n}^f_{\mathbf{k}}\rangle-\langle \tilde{n}^c_{\mathbf{k}+\mathbf{q}}\rangle)\,,
\end{align}
\begin{align}
\Gamma^{1}_{\mathbf{kq}\mathbf{k}_1\mathbf{k}_2}=&U^2(|\tilde{x}'_{\mathbf{k}}\tilde{y}_{\mathbf{k}_1,\mathbf{k}+\mathbf{q},\mathbf{k}_2}|^2\nonumber\\
&-\tilde{x}'_{\mathbf{k}}\tilde{x}'_{\mathbf{k}_1+\mathbf{k}_2-\mathbf{k}-\mathbf{q}}
\tilde{y}_{\mathbf{k}_1,\mathbf{k}+\mathbf{q},\mathbf{k}_2}\tilde{y}_{\mathbf{k}_1,\mathbf{k}_1+\mathbf{k}_2-\mathbf{k},\mathbf{k}_2})\nonumber\\
\times &\Big[\langle\tilde{n}^f_{\mathbf{k}}\rangle\langle
\tilde{n}^f_{\mathbf{k}_1+\mathbf{k}_2-\mathbf{k}-\mathbf{q}}\rangle(1-\langle\tilde{n}^c_{\mathbf{k}_1}\rangle
-\langle\tilde{n}^f_{\mathbf{k}_2}\rangle)\nonumber\\
&-\langle\tilde{n}^f_{\mathbf{k}_2}\rangle
\langle\tilde{n}^c_{\mathbf{k}_1}\rangle(1-\langle\tilde{n}^f_{\mathbf{k}}\rangle-
\langle\tilde{n}^f_{\mathbf{k}_1+\mathbf{k}_2-\mathbf{k}-\mathbf{q}}\rangle)\Big]\,,
\end{align}
\begin{align}
\Gamma^{2}_{\mathbf{kq}\mathbf{k}_1\mathbf{k}_2}
=&U^2(|\tilde{x}_{\mathbf{k}+\mathbf{q}}\tilde{y}'_{\mathbf{k}_1,\mathbf{k}_2,\mathbf{k}-\mathbf{k}_1+\mathbf{k}_2}|^2\nonumber\\
&-\tilde{x}_{\mathbf{k}+\mathbf{q}}
\tilde{x}_{\mathbf{k}_2-\mathbf{q}}\tilde{y}'_{\mathbf{k}_1\mathbf{k}_2,\mathbf{k}-\mathbf{k}_1+\mathbf{k}_2}\tilde{y}'_{\mathbf{k}_1,\mathbf{k}+\mathbf{q},\mathbf{k}-\mathbf{k}_1+\mathbf{k}_2})\nonumber\\
\times &
\Big[\langle\tilde{n}^c_{\mathbf{k}_2}\rangle\langle\tilde{n}^c_{\mathbf{k}+\mathbf{q}}\rangle
(1-\langle\tilde{n}^f_{\mathbf{k}-\mathbf{k}_1+\mathbf{k}_2}\rangle-\langle\tilde{n}^c_{\mathbf{k}_1}\rangle)\nonumber\\
&-\langle\tilde{n}^c_{\mathbf{k}_1}\rangle
\langle\tilde{n}^f_{\mathbf{k}-\mathbf{k}_1+\mathbf{k}_2}\rangle
(1-\langle\tilde{n}^c_{\mathbf{k}_2}\rangle-\langle\tilde{n}^c_{\mathbf{k}+\mathbf{q}}\rangle)\Big] \, .
\end{align}
Here the expectation values 
$\langle \tilde{n}_{\mathbf k}^c\rangle = \langle c_{\mathbf k}^\dagger  c^{}_{\mathbf k}\rangle_{\tilde{\mathcal{H}}}$ 
and 
$\langle \tilde{n}_{\mathbf k}^f\rangle =  \langle f_{\mathbf k}^\dagger  f^{}_{\mathbf k}\rangle_{\tilde{\mathcal{H}}}$,
are formed with the renormalized Hamiltonian $\tilde{\cal H}$ and  can easily be evaluated due to the 
diagonal form of  $\tilde{\mathcal H}$.  Note that 
the pole structure of the first (coherent) term of Eq.~\eqref{chi1} 
describes the continuum of particle-hole excitations.  Of course, we have 
$\omega_{\bf 0}({\bf 0})=E_g^{(0)}$, and in view of the form of the $c$ and $f$ band dispersions
$\omega_{\bf k}({\bf 0}) > \omega_{\bf 0}({\bf 0})\;\forall {\bf k}\neq {\bf 0}$.
Therefore the possibility of  ${\bf q}={\bf 0}$ excitations with positive energy 
indicates that the system is in the semiconducting regime. If one tries to determine
the semimetal-semiconductor boundary, i.e., $T_{MI}$, from the pole structure of $\chi({\bf q},\omega)$ 
this assertion is valid to leading order only; the second and third term of~\eqref{chi1} might lead to a shift of the 
lowest excitation energy in the ${\bf q}={\bf 0}$ sector. As shown below, this effect is negligible however:
the values of $T_{MI}$ derived from $E_g^{(0)}$ 
are in accord with the  results obtained 
from the dynamical susceptibility. 

The imaginary part of $\chi({\mathbf q},\omega)$ reads
\begin{align}\label{Imchi}
-\frac{1}{\pi}\textrm{Im}\chi(\mathbf{q},\omega)=&\frac{1}{N}\sum_{\mathbf{k}}\Gamma^{0}_{\mathbf{kq}}\delta[\omega-\omega^{}_{\mathbf k}(\mathbf q)]\nonumber\\
\hspace*{-2.3cm}&+\frac{1}{N^3}\sum_{\mathbf{k}\mathbf{k}_1\mathbf{k}_2}
[\Gamma^{1}_{\mathbf{kq}\mathbf{k}_1\mathbf{k}_2}\delta(\omega-E^{(1)}_{\mathbf{kq}\mathbf{k}_1\mathbf{k}_2})
\nonumber\\
&\hspace*{1.2cm}
-\Gamma^{2}_{\mathbf{kq}\mathbf{k}_1\mathbf{k}_2}
\delta(\omega-E^{(2)}_{\mathbf{kq}\mathbf{k}_1\mathbf{k}_2})],
\end{align}
where possible non-zero excitations outside the particle-hole continuum point to the existence of
excitonic resonances. 

\begin{figure}[t]
    \begin{center}
\includegraphics[angle = -0, width = 0.47\textwidth]{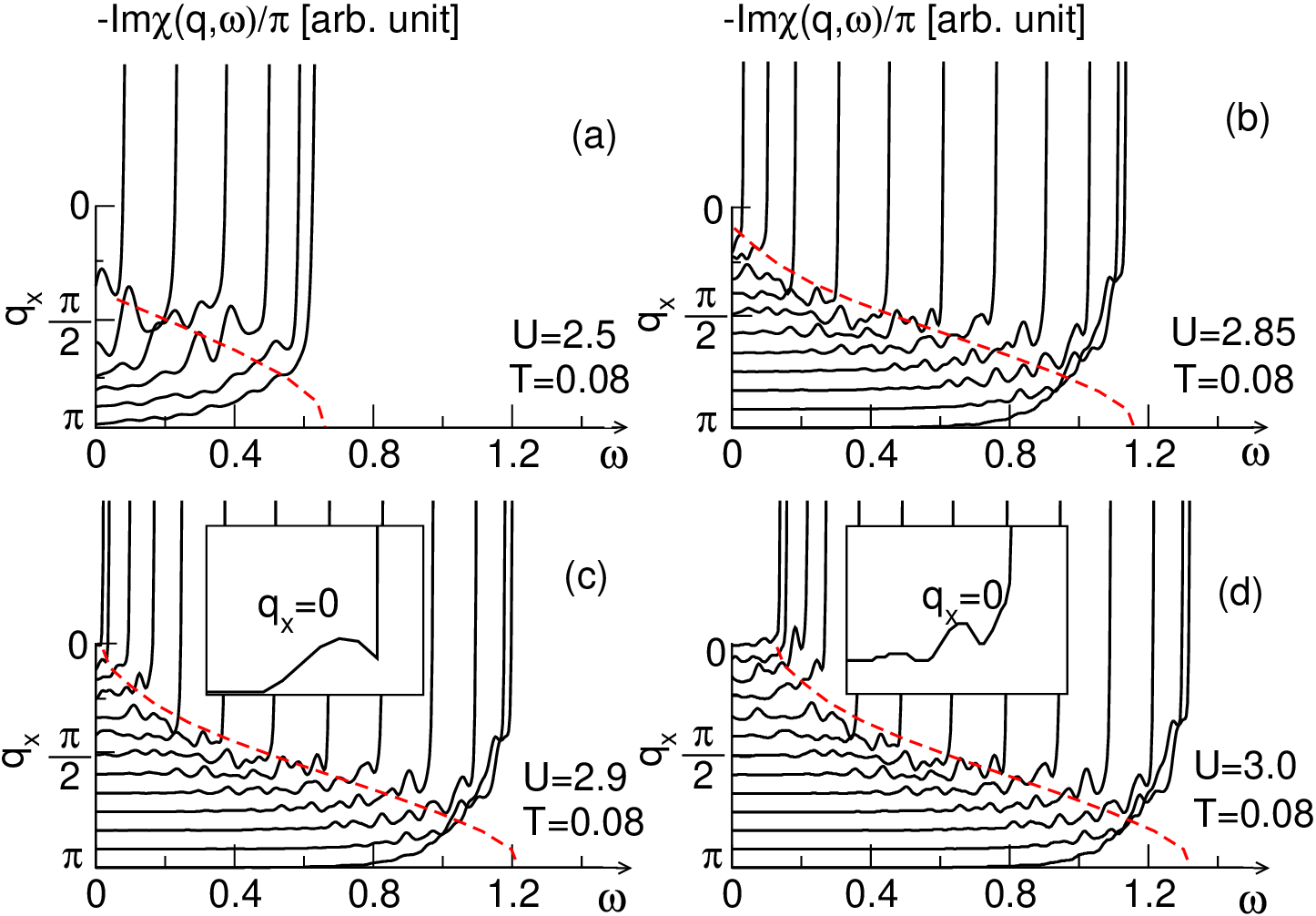}
\includegraphics[angle = -0, width = 0.47\textwidth]{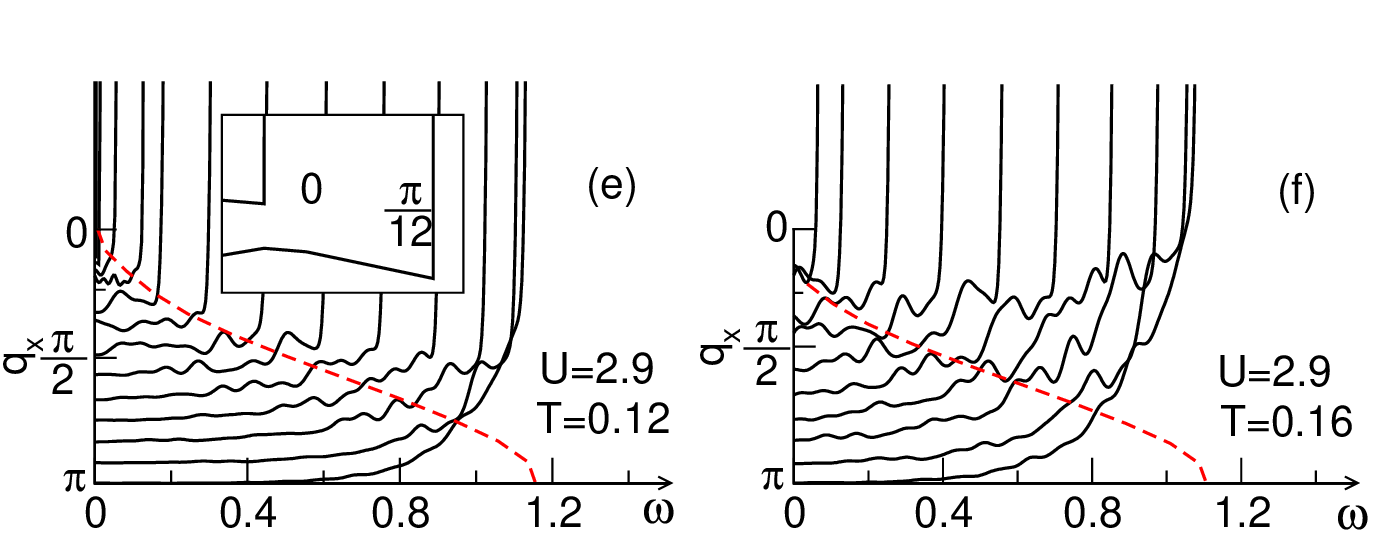}
    \end{center}

\caption{(Color online) Imaginary part of the dynamical 
susceptibility $-\textrm{Im}\chi(\mathbf{q},\omega)/\pi$ 
versus frequency vertically shifted to show its values for different
momenta along the $(q_x,0)$--direction (black solid lines). 
The boundaries to the particle-hole continuum 
are marked by red dashed lines. The insets magnify 
the low-frequency small-momentum region whenever a gap appears for 
the electron-hole excitations. Bare band parameters 
are $\varepsilon^f=-2.4$, $t_f=-0.3$.}
\label{fig:S00P0}
\end{figure}

Figure~\ref{fig:S00P0} diplays our numerical results for 
the imaginary part of the exitonic 
susceptibility  $-\textrm{Im}\chi(\mathbf{q},\omega+i0)/\pi$ as a 
function of $\omega$ for different momenta
$(q_x,0)$ between $q_x=0$ and $\pi$ (black solid lines), 
indicated by the scale attached to the ordinate axis. 
In all panels the red dashed line represents  the
boundaries to the particle-hole continuum  (indicated by the strong upturns). 
The small bumps in the figures correspond to excitonic resonances, where the maximum most closely located 
to the particle-hole continuum refers to an excitonic bound state.

In panels (a) to (d) the temperature is kept to $T=0.08$. 
For $U=2.5$ respectively $2.85$
(which are still larger than $U_{c1}$ however), 
$T>T_{MI}$, and the system is in the semimetallic region.  
No excitonic resonances  can be found for ${\bf q}=0$ in this case,
since $E_g^{(0)}$ is already negative (cf. fig.~\ref{fig:ED_T}).  
There is a significant increase of the spectral weight of 
the finite-${\bf q}$ excitonic resonances by going over from $U=2.5$ to $U=2.85$. 
In contrast, in panel (c), where $U=2.9$, we have $T_{EI} <T<T_{MI}$, 
and the system realizes a semiconductor (cf.~figs.~\ref{fig:PD} 
and \ref{fig:ED_T}). Now a weak excitonic resonance is found 
at momentum ${\mathbf q}=0$ which can more clearly be seen from the inset. 
For $U=3.0>U_{c2}$ [panel~(d)] the EI phase 
is not realized even for $T=0$ (cf.~fig.~\ref{fig:ED_T}), and again 
excitons can be formed for all values of $q_x$. 
Note that excitons with finite momentum ${\mathbf q}$ can be 
created in both semiconductor and semimetal  
cases [cf. panels (a) to (d)]. 
Their resonance positions follow from  eqs.~\eqref{er1} and \eqref{er2}. 

In the two lowermost panels (e) and (f) the Coulomb interaction 
is fixed to $U=2.9$. Increasing the temperature from $T=0.12$~(e) 
to $T=0.16$~(f) 
the system passes the semiconductor-semimetal transition. 
Although, at $T=0.12$, 
the system is very close to the transition point (cf. fig.~\ref{fig:ED_T}), 
excitons with zero momenta may form (see inset).  
In contrast,  no ${\mathbf q}=0$ excitons  can exist for the temperature considered in panel (f). Here we observe only excitonic resonances with 
finite momenta.  For the higher temperature case these    
resonances are more smeared out and their weight is enhanced. 
Therefore excitonic 
states with ${\bf q} \neq  0$ can easier be occupied for this case. 
  
Figure~\ref{fig:S00P0} clearly shows that it is 
possible to extract the temperature $T_{MI}$ just as well 
by monitoring the appearance of excitonic  resonances in 
the imaginary part of $\chi(\mathbf{q},\omega+i0)$ at ${\mathbf q}=0$.

\section{Conclusions}
In summary, we have performed a detailed investigation of the two-dimensional  extended  Falicov-Kimball model
by means of the projector-based renormalization method. Thereby we established the long-predicted existence of an intervening excitonic insulator phase at the semimetal-semiconductor transition below some critical $T_{EI}$ (see fig.~\ref{fig:PD}). 
We derived the renormalized quasiparticle band structure which shows a correlation-induced single-particle gap and $c$-$f$ electron coherence in the low-temperature EI state and reflects the metal-insulator transition at $T_{MI}$ for higher temperatures. Analyzing the imaginary part of the excitonic pair 
susceptibility, 
we demonstrate that on the semiconductor side of this phase transition, 
preformed excitons with zero momentum exist above $T_{EI}$. 
On the other hand, excitonic bound states (resonances) with finite momentum 
may appear on both---semiconducting and semimetallic---sides of 
the metal-insulator transition, but these excitons will not condense
for the studied direct band gap situation. We therefore corroborate the scenario, suggested by Bronold and Fehske~\cite{BF06}, that in the semiconducting region  the EI phase is surrounded by  an excitonic halo consisting of free electrons, holes and tightly bound zero-momentum excitons. Forming the EI state, 
the latter undergo a Bose-Einstein condensation state as the temperature is lowered.  
Contrariwise there is a well-defined (large) Fermi surface in the semimetallic  regime,
and the EI state can be envisaged as composed of BCS-type electron-hole pairs. 
\acknowledgements
The authors would like to thank F. X. Bronold, D. Ihle, H. Stolz, and B. Zenker 
for valuable discussions. HF acknowledges a Gordon Godfrey 
fellowship by the UNSW, where this work was completed.  
Research was supported by the DFG through SFB 652.

\end{document}